\documentclass[aps,12pt,amssymb,preprintnumbers,amsmath,amsfonts]{revtex4}
\usepackage{latexsym}
\usepackage[latin1]{inputenc}
\usepackage{amsmath}
\newcommand{\ba}{\begin{eqnarray}}
\newcommand{\ea}{\end{eqnarray}}

\newcommand{\be}{\begin{equation}}
\newcommand{\ee}{\end{equation}}
\newcommand{\pa}{\partial}

\newcommand{\Tr}{{\rm Tr}}
\newcommand{\vt}{{{\tilde{v}}}}
\newcommand{\vts}{{\slashed{\tilde{v}}}}
\newcommand{\Eq}[1]{Eq.~\eqref{#1}}
\newcommand{\aq}{{\vert {\bf{q}} \vert}}

\usepackage{hyperref}  

\usepackage{dcolumn}
\usepackage[dvips]{graphicx}
\usepackage{color}
\usepackage{slashed}

\usepackage{ulem}
\definecolor{stcol}{rgb}{1,0,1}

\begin{document}
                                                                                
\date{\today}

\title{Consistent relativistic chiral kinetic theory: A derivation from on-shell effective theory}

\author{Stefano Carignano}
\email{carignano@lngs.infn.it}
\affiliation{Instituto de Ciencias del Espacio (ICE, CSIC) \\
C. Can Magrans s.n., 08193 Cerdanyola del Vall\`es, Catalonia, Spain
and \\
 Institut d'Estudis Espacials de Catalunya (IEEC) \\
 C. Gran Capit\`a 2-4, Ed. Nexus, 08034 Barcelona, Spain
}
\author{Cristina Manuel}
\email{cmanuel@ice.csic.es}
\affiliation{Instituto de Ciencias del Espacio (ICE, CSIC) \\
C. Can Magrans s.n., 08193 Cerdanyola del Vall\`es, Catalonia, Spain
and \\
 Institut d'Estudis Espacials de Catalunya (IEEC) \\
 C. Gran Capit\`a 2-4, Ed. Nexus, 08034 Barcelona, Spain
}
\author{Juan M. Torres-Rincon}
\email{juan.torres-rincon@stonybrook.edu}
\affiliation{Department of Physics and Astronomy, Stony Brook University, Stony Brook, New York 11794-3800, USA.}

\begin{abstract}
We formulate the on-shell effective field theory (OSEFT) in an arbitrary frame and study its reparametrization invariance, which
ensures that it respects  Lorentz symmetry.
In this formulation the OSEFT Lagrangian looks formally equivalent to the sum over lightlike velocities
of soft collinear effective field theory in the Abelian limit, but differences remain in the scale of the gauge fields involved in the two effective theories.
We then use  the OSEFT Lagrangian expanded in inverse powers of the on-shell energy to derive how the classical transport equations for charged massless
fermions are corrected by quantum effects, as derived from quantum field theory. We provide a formulation in  a full covariant way and explain how the consistent form of the chiral
anomaly equation can be recovered from our results. We also show how the side-jump transformation of the distribution function associated with
massless charged fermions can be derived from the reparametrization invariance transformation rules of the OSEFT quantum fields.
Finally, we discuss the differences in our results with  respect to others found in the literature.
\end{abstract}
\maketitle

\section{Introduction}

In this paper we use the so-called on-shell effective field theory (OSEFT) \cite{Manuel:2014dza,Manuel:2016wqs,Manuel:2016cit} to provide a derivation of the transport equations obeyed by charged chiral fermions
beyond the classical limit approximation. 

A formulation of transport theory for chiral fermions  has been developed in  Refs.~\cite{Son:2012wh,Stephanov:2012ki,Son:2012zy,Chen:2012ca}, starting with  the
action of a point particle modified by the Berry curvature, together with a modified Poisson bracket structure. Other alternative approaches to derive
the same transport equation can be found in the literature \cite{Manuel:2013zaa,Manuel:2014dza,Hidaka:2016yjf,Hidaka:2017auj,Mueller:2017lzw,Mueller:2017arw,Hidaka:2018ekt,Gorbar:2016ygi,Gorbar:2017awz,Huang:2018wdl,Gao:2018wmr}.

The first derivation of chiral kinetic theory (CKT) from quantum field theory was made in Ref.~\cite{Son:2012zy} for systems at finite density and zero temperature,
using the so-called high density effective field theory (HDET) \cite{Hong:1998tn}. OSEFT was actually proposed to provide a similar derivation that could be valid also in a thermal background, where antifermions are also
relevant degrees of freedom. 
Regardless of the background,  transport equations describe the propagation of an on-shell quasiparticles, and therefore it seems natural to use  for their derivation 
an effective field theory approach that describes only the propagation of on-shell degrees of freedom, as OSEFT, while off-shell modes are integrated out. 
Let us stress that the notion of on-shell quasiparticle  depends on the energy scales one is looking at in the
system under consideration.  
It is well known that for plasmas at finite temperature $T$ only the high energy modes of order $T$ can be considered as quasiparticles and their evolution studied with classical transport equations \cite{Blaizot:2001nr,Litim:2001db,Kelly:1994ig}, while the same picture does not apply to lower energy modes. To get corrections to the classical point-particle picture  described above from quantum field theory, one simply has to study
how the off-shell modes modify the evolution of the highly energetic modes. These corrections are taken into account in the OSEFT Lagrangian, and expressed as operators of increasing dimension over powers of the on-shell energy scale,
so that these modifications can be described with the accuracy one desires. The OSEFT Lagrangian can then be used to derive how the classical transport picture is modified, by using for their derivation an increasing number
of terms in the high energy expansion.

One of the advantages of our formulation is that it may allow us to derive transport equations in full covariant form, and derive their properties under Lorentz transformations.
While the initial proposals of CKT were not given in a covariant form, it was soon realized that it would have peculiar properties under Lorentz transformations~\cite{Chen:2014cla,Chen:2015gta},
especially seen when formulating two-body collisions, but also expressed in the so-called side-jump behavior of the distribution function of CKT, that expresses that it is frame dependent.

We present in this paper a derivation of CKT in a covariant way, as derived from OSEFT, and explain how the side-jump effects can be deduced from the same symmetries of that
effective field theory. While previous formulations of OSEFT were given in the preferred frame of the thermal bath, we generalize it to an arbitrary frame, introducing the frame vector $u^\mu$. The resulting OSEFT Lagrangian then looks formally equivalent to that corresponding to a sum over velocities of the so-called soft collinear effective field theory (SCET) \cite{Bauer:2000yr,Bauer:2001yt,Beneke:2002ph,StewartLects}, although there are some differences, as will be discussed in the following. 
We further study  the reparametrization invariance of OSEFT, that ensures that our
formalism is respectful of Lorentz invariance.

We compute both the vector current and axial current in the OSEFT, by taking functional derivatives to the action, and take these expressions to deduce the corresponding values in the transport framework, which
requires a Wigner transformation of a two-point function, together with a gradient expansion. As very clearly explained in the review \cite{Landsteiner:2016led}, such a definition can only lead to the consistent
form of the chiral anomaly, rather than the covariant form. We check from our expressions that this is indeed the case.

Our final form of the relativistic chiral transport equation mainly differs from that introduced in Refs.~\cite{Hidaka:2016yjf,Hidaka:2017auj,Hidaka:2018ekt}, in pieces that may be subleading when considering
effects close to thermal equilibrium, but that might be relevant for studies off equilibrium, and also in the gradient terms of the gauge fields. It also differs, when fixing the frame, with the chiral transport equation obtained from the modified form of the  one-point particle action.

Our paper is organized as follows. In Sec.~\ref{OSEFT=SumSCET}, we formulate OSEFT in an arbitrary frame, introducing a frame vector and showing its formal
equivalence with soft collinear effective field theory. In Sec.~\ref{RiOSEFT} we study the reparametrization invariance of this effective field theory, a basic ingredient to show
that it is respectful of Lorentz symmetry. In Sec.~\ref{WignerSection}, we introduce the basic two-point function in the OSEFT that will be used 
to derive the basic set of transport equations. The main content of the paper is in Sec.~\ref{Derivation}, with the derivation of the collisionless transport equation, first using
the OSEFT variables in Sec.~\ref{OSEFTcomputation}, and then expressed in terms of the QED original variables in Sec.~\ref{backwards}. In Sec.~\ref{Consistentcurrent+anomaly},
we derive both the vector and axial current obtained in the OSEFT approach, and check that they obey the consistent form of the quantum anomalies. In Sec.~\ref{Sidejumps=RI}
we derived the side-jump transformation of the distribution function from the reparametrization invariance transformations of the OSEFT quantum fields. We conclude in 
Sec.~\ref{discussion}, where we summarize our main findings, and give a possible interpretation of the origin of the discrepancy of our results with alternative approaches. In Appendix \ref{compute-Is}
we give some details of our computations, while in Appendix \ref{CME-app} we show how to obtain the chiral magnetic effect from our formulation.

We use natural units $\hbar = c = k_B =1$ and metric conventions $g^{\mu \nu } =(1,-1,-1,-1)$. We also use boldface letters to denote 3-vectors.

\section{OSEFT in an arbitrary frame and SCET}
\label{OSEFT=SumSCET}

Let us review the OSEFT as originally formulated~\cite{Manuel:2014dza,Manuel:2016wqs}, introducing the basic fields and notation.
Let us recall that the propagation of an on-shell massless fermion is  described by its energy $p$, with $p > 0$, and the lightlike 4-velocity $v^\mu= (1, {\bf v})$, where
${\bf v}$ is three-dimensional unit vector, and thus its 4-momentum is $p^\mu = p v^\mu$. However, for a fermion close to being on-shell, its 4-momentum can be expressed as 
\be
q^\mu = p v^\mu + k^\mu \ ,\label{eq:Qpart}
\ee
where $k^\mu$ is the residual momentum ($k_\mu \ll p)$, {\it i.e}. the part of the momentum which makes $q^\mu$ off shell. 
 A similar decomposition of the momentum for almost on-shell antifermions can be done as follows,
\begin{equation}
q^{\mu}=-p\tilde{v}^{\mu}+k^{\mu}\ ,\label{eq:QAntiPart}
\end{equation}
where  $\tilde{v}^{\mu}=(1,-\bf{v})$~.

The Dirac field  can be written as
\begin{equation}
\psi_{ v, \tilde v}=e^{-ipv\cdot x}\left( P_v \chi_{v}(x) +  P_{\tilde v} H_{\tilde v}^{(1)}(x)\right)+e^{ip\tilde{v}\cdot x} \left( P_{\tilde v} \xi_{\tilde v}(x)+P_v   H_{v}^{(2)}(x) \right)\ ,
\label{eq:Fields}
\end{equation}
where  the basic OSEFT quantum fields  obey
\begin{eqnarray}
&P_v \chi_v = \chi_v &\ , \qquad P_{\tilde v} \chi_v = 0 \ , \\
&P_{\tilde v} \xi_{\tilde v} = \xi_{\tilde v} & \ , \qquad P_v \xi_{\tilde v} = 0 \ ,
\end{eqnarray}
and the particle/antiparticle projectors are expressed as
\begin{equation}
P_v  =  \frac{1}{2}\gamma\cdot v \ \gamma_{0} \ , \qquad 
P_{\tilde v}  =  \frac{1}{2}\gamma\cdot\tilde{v}\ \gamma_{0} \ . \label{eq:AProj}
\end{equation}

It is possible to integrate out the $H^{(1,2)}$ fields of the QED Lagrangian \cite{Manuel:2014dza}, to have an effective theory
for the fields $\chi_v$ and $\xi_{\tilde v}$ only. 

If we assume that the physical phenomena we aim to describe are dominated by the contribution of on-shell particles,
then the corresponding OSEFT Lagrangian can  be written as a sum over the different values of the on-shell momenta as
\begin{equation}
\label{sum-OSEFTL}
\mathcal{L}=\sum_{p, {\bf v}}\mathcal{L}_{p, {\bf v}}\,,
\end{equation}
where the precise meaning of the sum displayed in Eq.~(\ref{sum-OSEFTL}) is not needed at this stage (we will come back to this point later on; see also Ref.~\cite{Manuel:2016wqs}), and
\begin{eqnarray}
\mathcal{L}_{p,{\bf v}}\ & = & \mathcal{L}_{p, v} + \widetilde {\mathcal{L}}_{p,{\tilde v}} 
\nonumber \\
& = &
\bar \chi_{v}(x) \left(i\, v\cdot D\,
+i \slashed{D}_{\perp} \frac{1}{2 p + i \tilde{v}\cdot D   }\,i \slashed{D}_{\perp}
\right) \gamma^0 \chi_{v}(x) \nonumber \\
&+&{\bar \xi}_{\tilde v}(x)  \left(i\, \tilde{v}\cdot D\,
+i \slashed{D}_{\perp} \frac{1}{ - 2 p + i  v\cdot D   }\,i \slashed{D}_{\perp} \right)\gamma^0 \xi_{\tilde v}(x) \ ,
\label{eq:Leff}
\end{eqnarray}
where  $D_\mu = \pa_\mu + i e A_\mu$ is the covariant derivative, $\slashed{D}_{\perp} = P^{\mu \nu}_{\perp} \gamma_\mu D_\nu$
and
\be
\label{transverse-projector}
P^{\mu \nu}_{\perp} = g^{\mu \nu} - \frac 12 \left( v^\mu {\tilde v}^\nu +v^\nu {\tilde v}^\mu\right) \ ,
\ee
is minus the transverse projector to ${\bf v}$, written in covariant form. Note that with our conventions $k^2_\perp = P^{\mu \nu}_{\perp} k_\mu k_\nu = - {\bf k}^2_\perp$.
From now on, and as done in Ref.~\cite{Manuel:2016wqs}, whenever we write  a tensor with the symbol $\perp$, it means that a transverse projector applies to all its Lorentz indices.
If only the transverse projector is applied to one of the indices, we will write $\perp$ only affecting that index. Thus, $\sigma_\perp ^{\mu \nu} = P^{\mu \alpha}_\perp P^{\nu \beta}_\perp \sigma_{\alpha \beta}$,
while  $\sigma^{\mu_\perp \nu} = P^{\mu \alpha}_\perp  g^{\nu \beta} \sigma_{\alpha \beta}$.

In the original formulation of the OSEFT  a choice of frame was made 
\cite{Manuel:2014dza,Manuel:2016wqs}.  The energies of the on-shell particles in Eq.~(\ref{eq:Qpart}) are measured in the same frame where, for example, the thermal
bath is defined. If we want to express the same OSEFT Lagrangian in an arbitrary frame, we will then have to introduce a timelike vector $u^\mu$
which defines that frame.
Then one could write all the above different equations simply by replacing
\be 
p  \rightarrow u^\mu p_\mu  \equiv E \ , \qquad   \gamma_0   \rightarrow \gamma_\mu u^\mu \ .
\ee

With our specific choice of variables $v^\mu$ and ${\tilde v}^\mu$, then it is not difficult to see that
\be
\label{frame-vel}
u^\mu = \frac{v^\mu +{\tilde v}^\mu}{2} \ .
\ee 
Note that in OSEFT  $u^\mu$ is not an independent vector, once $v^\mu$ and ${\tilde v}^\mu$ have been defined. 
While in the static frame we chose a particular  definition of the vectors $v^\mu$ and ${\tilde v}^\mu$, which implicitly assumed that
$u^\mu = (1,0,0,0)$, in an arbitrary frame we will only ask that
these lightlike vectors obey
\be
\label{light-condition}
v^2 = {\tilde v}^2 = 0 \ , \qquad  v \cdot {\tilde v} =2 \ .
\ee
Thus, $u \cdot v =1$ and $u^2 =1$ are automatically fulfilled.

In our formulation of the OSEFT in an arbitrary frame, we will sometimes use ${\tilde v}^\mu$, and sometimes we will use $u^\mu$. 
The last option is convenient, as in kinetic theory it may appear also in the thermal equilibrium distribution  associated with the massless particles.

As for the  the particle/antiparticle projectors in an arbitrary frame, we will write them as
\begin{eqnarray}
P_v & = & \frac{1}{2}  \slashed{v} \, \slashed{u} =  \frac{1}{4}  \slashed{v} \,  \slashed{\tilde v} 
\, \label{eq:AProj-bis1}\\
P_{\tilde v} & = & \frac{1}{2}  \slashed{\tilde v} \, \slashed{u}  = \frac{1}{4}   \slashed{\tilde v} \,\slashed{v}     \ ,
\label{eq:AProj-bis2}
\end{eqnarray}
where we used that  $\slashed{v} \,\slashed{v} =  \slashed{\tilde v} \, \slashed{\tilde v} = 0$.

The OSEFT Lagrangian in a general frame is then written down as 
\be
\mathcal{L} = \sum_{E, v} \left( \mathcal{L}_{E, v}+ \mathcal{L}_{-E, {\tilde v} }\right) \ ,
\ee
where
\begin{eqnarray}
\mathcal{L}_{E, v}+ \mathcal{L}_{-E, {\tilde v}}\ & = &
\bar \chi_{v}(x) \left(i\, v\cdot D\,
+i \slashed{D}_{\perp} \frac{1}{2 E + i \tilde{v}\cdot D   }\,i \slashed{D}_{\perp}
\right)  \frac{ \slashed{\tilde v} }{2}  \chi_{v}(x) \nonumber \\
&+&{\bar \xi}_{\tilde v}(x)  \left(i\, \tilde{v}\cdot D\,
+i \slashed{D}_{\perp} \frac{1}{ - 2 E + i  v\cdot D   }\,i \slashed{D}_{\perp} \right) \frac{ \slashed{v} }{2} 
 \xi_{\tilde v}(x)
\,.\label{Leff-SCET}
\end{eqnarray}
where we have used that 
$$  \slashed{v} \chi_v = 0 \ , \qquad    \slashed{\tilde v} \xi_{\tilde v}=0 \ .  $$

It is noteworthy that Eq.(\ref{Leff-SCET}) formally looks similar to  the Lagrangian of soft-collinear effective field theory \cite{Bauer:2000yr,Bauer:2001yt,Beneke:2002ph,StewartLects}. The corresponding projectors
 Eqs.~(\ref{eq:AProj-bis1}, \ref{eq:AProj-bis2}) are also those used in SCET.
We note that the explicit forms of the OSEFT and SCET Lagrangians differ  because of our
different convention in defining the quantum fluctuating fields: in SCET, the exponential terms of Eq.~(\ref{eq:Fields}) have been included
in the quantum fields of the effective  theory. We also explicitly separate the contribution of particles and antiparticles. Further, we recall that we are considering
an effective field theory for QED, while SCET is an effective field theory for QCD.

After noticing the  above formal similarities of SCET and OSEFT when the latter is formulated in an arbitrary frame,  it has to be stressed that they are still different effective field theories. SCET was
originally formulated to describe the physics associated with highly energetic jets
in vacuum, and there are only two lightlike vectors in the theory, $v^\mu$ and ${\tilde v}^\mu$,  fixed by the direction of the jet. In SCET, the covariant derivatives are associated with collinear and ultrasoft gauge fields.  OSEFT was in principle developed to describe
many body particle systems, close to thermal equilibrium, where one can consider having many on-shell particles and their propagation in the background of soft gauge fields. Thus, for a fixed value of the energy 
there might be particles moving in all arbitrary (lightlike) directions, and a sum over $v^\mu$ is displayed in the final Lagrangian, which is absent in SCET. 
In OSEFT, the covariant derivatives we use mainly contain soft gauge fields.

OSEFT  also uses a different notation, which makes clear that its main goal is to make  an analytical
expansion in powers of the inverse of the on-shell energy $1/E$. At finite temperature and/or density we will obtain different expressions multiplied by
a particle distribution function. After integration over momenta, this expansion on the inverse of the on-shell energy  will turn out to give an expansion in powers of the inverse
of the temperature and/or chemical potential \cite{Manuel:2016wqs,Manuel:2016cit}.

After mentioning the explicit similarities and differences of these two effective field theories, it is possible to use some of the results obtained in SCET to learn
about some properties of OSEFT, such as that of reparametrization invariance, which will be discussed in the following section.

\section{Reparametrization invariance of OSEFT}
\label{RiOSEFT}

Reparametrization invariance (RI) is the symmetry associated with the ambiguity of the decomposition of
the momentum $q^\mu$ performed in Eq.~(\ref{eq:Qpart}). If $M^{\mu \nu}$ defines the six Lorentz generators of $SO(3,1)$, the decomposition of Eq.~(\ref{eq:Qpart}) suggests
an apparent breaking of five Lorentz generators, namely, $\{ v_\mu M^{\mu \nu}, u_\mu M^{\mu \nu} \}$, or, equivalently, $\{ v_\mu M^{\mu \nu}, {\tilde v}_\mu M^{\mu \nu} \}$.
However, it is possible to show that the OSEFT Lagrangian is RI invariant, which is equivalent to saying that is Lorentz invariant. Let us stress that this reduces to the study of the RI of SCET for
every sector of the theory defined by the vectors $v_\mu$ and ${\tilde v}_\mu$, something which has been extensively investigated \cite{Manohar:2002fd}. The fact that the covariant derivatives
displayed in SCET and OSEFT contain gauge fields of different scales does not, however, affect the proof of RI, which turns out to be formally equivalent in the two effective field theories.

Let us review how this effectively works. The Dirac field defined in Eq.~(\ref{eq:Fields}) should be the same independent of the choice of the parameters used to define the effective field theory; thus,
\be
\label{eq-RI}
\psi_{ v, \tilde v} (x)= \psi'_{ v', \tilde v'}(x) \ .
\ee

As in SCET, we will see that the effective field theory action  remains invariant under  infinitesimal changes of the vectors $v^\mu$ and ${\tilde v}^\mu$ that preserve their basic properties 
expressed in Eq.~(\ref{light-condition}). It is possible to show that the OSEFT Lagrangian is invariant under the following symmetries
\begin{eqnarray}
({\rm I}) \, \Bigg \{ \begin{array}{ccl}
v^\mu &\rightarrow &v^\mu + \lambda_\perp^\mu  \\
{\tilde v}^\mu & \rightarrow &{\tilde v}^\mu  
\end{array}   \qquad  ({\rm II}) \,  \Bigg \{ \begin{array}{ccl}
v^\mu &\rightarrow &v^\mu   \\
{\tilde v}^\mu & \rightarrow &{\tilde v}^\mu + \epsilon_\perp^\mu 
\end{array} 
  \qquad 
({\rm III}) \, \Bigg \{ \begin{array}{ccl}
v^\mu &\rightarrow & (1+ \alpha) v^\mu  \\
{\tilde v}^\mu & \rightarrow & (1-\alpha){\tilde v}^\mu 
  \end{array}  
\end{eqnarray}
where $\{ \lambda_\perp^\mu, \epsilon_\perp^\mu, \alpha\}$ are five infinitesimal parameters, and
$v \cdot \lambda_\perp = v \cdot \epsilon_\perp = {\tilde v} \cdot \lambda_\perp = {\tilde v} \cdot \epsilon_\perp =0$.
Please note that  the transformation rule of the vector $u^\mu$ can be deduced from 
Eq.~(\ref{frame-vel}).

Just to have a flavor of the meaning of the above symmetries, let us imagine one fixes the values of the two lightlike vectors as $v^\mu = (1, 0,0,1)$ and ${\tilde v}^\mu = (1, 0,0, -1)$. Then, apparently, there are five broken generators in the OSEFT, 
which are $Q_1^{\pm} = J_1 \pm K_2$, $Q_2^{\pm} = J_2 \pm K_1$, and $K_3$, where $J_i$  and $K_i$ are the generators of rotations and  boosts, respectively.
Then, type I refers to the combined action of an infinitesimal boost in the $x (y)$ direction and a rotation around the  $ y (x)$ axis, such that 
${\tilde v}^\mu$ is left invariant, with generators $(Q_1^-, Q_2^+)$.   Type II transformations are similar
but $(Q_1^+, Q_2^-)$ leave $v^\mu$ invariant, while type III is a boost along the direction 3, $K_3$.

It is also worth it to note that the the generators $(Q_1^+, Q_2^-,J_3)$  obey the $SE(2)$ Lie algebra, that is the symmetry group of the two-dimensional Euclidean plane. They correspond to what is known
as the Wigner little group associated with the vector $p^\mu =p v^\mu$  
 \cite{Stone:2015kla}, see also Refs.~\cite{Baskal,Duval:2014cfa,Duval:2014ppa}. Similarly, the generators $(Q_1^-, Q_2^+,J_3)$ correspond to Wigner's little group associated with $p^\mu = -p \tilde{v}^\mu$ (antiparticles). As discussed in Ref.~\cite{Stone:2015kla}  these Wigner translations are associated with shifts of the trajectory of finite wave packets of massless 
particles proportional to the particle's helicity.  

It is possible to check easily that our Lagrangian is invariant under the above three RI transformations \cite{Manohar:2002fd}, which formally
is equivalent to say that it is Lorentz invariant. Let us discuss these briefly, as they are the same RI symmetries of SCET. We will mainly focus now on what our different notation implies. We will concentrate in the following in the particle sector, as for antiparticles things
works analogously after trivial changes (namely, $u \cdot p \rightarrow - u \cdot p$ and $v^\mu \leftrightarrow {\tilde v}^\mu$).  We will also see  that the type ${\rm II}$ symmetry will allow us to generate the side-jumps that were discussed in the framework of chiral kinetic theory in  Ref.~\cite{Chen:2014cla}. This point will be discussed in Sec.~\ref{Sidejumps=RI}.

Let us first start with type I symmetry. The change in the vector $v^\mu$ implies a relabeling of what is called on-shell and residual parts
of the momentum defined in Eq.~(\ref{eq:Qpart}).
After a type I symmetry the on-shell part and residual  momenta change as
\ba
( u \cdot p) v^\mu & \rightarrow &( u \cdot p) v^\mu + \frac 12 (\lambda_\perp \cdot p) v^\mu + ( u \cdot p) \lambda_\perp^\mu \ ,
\\
k^\mu &\rightarrow & k^\mu - \frac 12 (\lambda_\perp \cdot p) v^\mu - ( u \cdot p) \lambda_\perp^\mu \ ,
\ea
respectively.
This implies that under a type I transformation the covariant derivatives acting on the fluctuating fields also transform.

Type II  symmetry implies that the new on-shell and residual momenta change as
\ba
( u \cdot p) v^\mu & \rightarrow & (u \cdot p) v^\mu + \frac12 (p \cdot \epsilon_\perp) v^\mu \ , \\
k^\mu & \rightarrow & k^\mu -   \frac12 (p \cdot \epsilon_\perp) v^\mu \ ,
\ea
while the type III transformation leads to the changes
\ba
( u \cdot p) v^\mu & \rightarrow &  ( u \cdot p) v^\mu (1 + 2 \alpha) - \alpha ({\tilde v} \cdot p) v^\mu \ , \\ 
k^\mu &\rightarrow & k^\mu - 2 \alpha E v^\mu+\alpha ({\tilde v} \cdot p) v^\mu \ , 
\ea
in the on-shell and residual momenta, respectively.

In Table~\ref{RI-transfor}, we summarize the transformation rules under all three types of transformations.

\begin{table}[h]
\begin{centering}
\resizebox{\textwidth}{!}{%
\begin{tabular}{|c || l | l | l |}
\hline
    &  Type I &  Type II & Type III \\  
 \hline
$v^\mu $                  & $v^\mu + \lambda_\perp^\mu $                   & $v^\mu$                                                              &  $v^\mu (1+\alpha)$                 \\
${\tilde v}^\mu$        & ${\tilde v}^\mu$                                            & ${\tilde v}^\mu + \epsilon_\perp^\mu$                 & $ {\tilde v}^\mu  (1-\alpha) $     \\
$u^\mu$                   & $u^\mu + \frac{\lambda_\perp^\mu}{2}$       & $u^\mu + \frac{\epsilon_\perp^\mu}{2}$             & $u^\mu (1-\alpha)+ \alpha v^\mu$ \\ 
$E $  	                & $E + \frac12 \lambda^\perp \cdot p$           &$ E + \frac12 (\epsilon_\perp \cdot p )$                & $ E (1+\alpha)- \alpha ({\tilde v} \cdot p) $ \\ 
$D_\mu$                  &$  D_\mu + i E \lambda_\mu^\perp + \frac i2 (\lambda_\perp \cdot p) v_\mu $
                                 & $ D_\mu+ \frac i2 (\epsilon_\perp \cdot p) v_\mu$    
                                 & $ D_\mu+ 2 i \alpha E\,  v_\mu  -i \alpha ({\tilde v} \cdot p) v_\mu  $   \\
$(v \cdot D) $           & $ (v \cdot D) + \lambda^\perp \cdot D^\perp $    &  $(v \cdot D)$                                                   & $ (v \cdot D) (1+\alpha) $ \\
$({\tilde v} \cdot D)$ & $({\tilde v} \cdot D) + i \lambda^\perp \cdot p  $     
                                 & $({\tilde v} \cdot D) + i \,\epsilon_\perp \cdot p + \epsilon_\perp  \cdot D_\perp $
                                 & $ ({\tilde v} \cdot D)  (1-\alpha)  + 4 i E \alpha - 2 i \alpha ({\tilde v} \cdot p) $    \\
$D_\mu^\perp$        & $ D_\mu^\perp - \frac{\lambda^\perp_\mu}{2} ({\tilde v} \cdot D)  - \frac{\tilde v_\mu}{2} \lambda^\perp \cdot D^\perp + i E \lambda_\mu^\perp $
                                 & $D_\mu^\perp - \frac{\epsilon^\perp_\mu}{2} ( v \cdot D)  - \frac{ v_\mu}{2} \epsilon^\perp \cdot D^\perp  $ 
                                 & $ D_\mu^\perp $    \\
$P_{v} $                   & $P_v + \frac 14 \slashed{\lambda}_\perp \slashed{\tilde v} $  & $ P_v - \frac 14 \slashed{\epsilon}_\perp \slashed{ v}   $ & $ P_v  $  \\
$\chi_{v} (x) $          & $ \left(1 +  \frac 14 \slashed{\lambda}_\perp \slashed{\tilde v} \right) \chi_{v} (x) $
                                 & $\left(1 +  \frac 12 \slashed{\epsilon}_\perp  \frac{1}{2E + i {\tilde v} \cdot D}  i \slashed{D}_\perp \right) \chi_{v} (x) $
                                 & $ \chi_{v} (x)$ \\
 \hline
 \end{tabular}
 }
\caption{Transformation rules in OSEFT under RI transformations of types I, II and III \label{RI-transfor}.}
\end{centering}
\end{table}

The OSEFT Lagrangian is invariant under these three  RI transformations:
  \cite{Manohar:2002fd} 
\be
\label{RI-inv-L}
\delta_{({\rm I})} \mathcal{L}_{E, v} = \delta_{({\rm II})} \mathcal{L}_{E, v}  = \delta_{({\rm III})} \mathcal{L}_{E, v} =  0 \ .
\ee

In explicit computations of Feynman diagrams, or derivations
of transport equations, we will expand the Lagrangian in power series of $1/E$. While Eq.~(\ref{RI-inv-L}) is exact to all orders in
a $1/E$ expansion, in a perturbative analysis in $1/E$ it is important to note that
 RI invariance implies that
different terms in the expansion are connected by symmetry. This comes from the fact that the covariant derivatives, or the fields, transform with terms
proportional to $E$.

For completeness, we will also mention other discrete symmetries of the OSEFT. Under parity, charge conjugation and time reversal,
the basic OSEFT fields transform as

\ba
&\chi_v(x) \rightarrow \gamma_0 \chi_{\tilde v}({\tilde x}_P) &\ , \qquad \xi_{\tilde v}(x) \rightarrow   \gamma_0 \xi_v ({\tilde x}_P)
\\
&\chi_v(x) \rightarrow -i \gamma^2 \xi^*_v( x) & \ , \qquad
\xi_{\tilde v}(x)  \rightarrow  -i \gamma^2 \chi^*_{\tilde v} (x) \\
&\chi_v(x) \rightarrow  - \gamma^1 \gamma^3 \chi_{\tilde v}(-{\tilde x}_T) & \ , \qquad
\xi_{\tilde v}(x) \rightarrow   - \gamma^1 \gamma^3 \xi_v (-{\tilde x}_T)
\ea
respectively, where if $x^\mu =(x_0, {\bf x})$, then ${\tilde x}_P^\mu =(x_0, - {\bf x})$, and  ${\tilde x}_T^\mu =(- x_0, {\bf x})$.

There is also a spin symmetry, which is not a $SU(2)$ symmetry but a $U(1)$ symmetry, which corresponds to helicity \cite{Baskal}.

\section{Wigner function in the OSEFT}
\label{WignerSection}

We focus our attention here on the basic Wigner function used in the following part of the paper for the derivation of the transport
equations from OSEFT.  We will use the Keldysh-Schwinger formulation,  allowing the time variables
 to take complex values, and define the two-point Green's functions of the OSEFT on the closed time-path contour. These are
 represented by a $2 \times 2$ matrix
\be
\label{progaKS}
S_{E,v} (x,y) = 
\left( \begin{array}{cc}
  S^{c}_{E,v} (x,y) & S^<_{E,v} (x,y) \\
 S^>_{E,v}(x,y) & S^{a} _{E,v}(x,y) \\
\end{array} 
\right ) =
\left( \begin{array}{cc}
\langle T  \chi_v(x) \bar \chi_v(y) \rangle & - \langle \bar \chi_v(y)  \chi_v(x) \rangle \\
\langle   \chi_v(x) \bar \chi_v(y) \rangle & \langle {\tilde T}  \chi_v(x) \bar \chi_v(y) \rangle 
\end{array} 
\right ) \ ,
\ee
where $T$ denotes time ordering and ${\tilde T}$ denotes anti-time ordering.

We will focus on one of the entries only, namely, $S^<_{E,v}$,
as this two-point function  depends only on  medium effects, while 
the diagonal entries of Eq.~(\ref{progaKS}) do also contain vacuum contributions. We will drop the superindex $^<$ in what follows to make  the notation lighter.

A similar two-point function can be introduced for the antiparticle quantum fluctuations.
From now on we will
focus on the particle's sector,  as the antiparticle's transport equations
may be derived similarly, and only involve some few changes to the particle's derivation ($E\rightarrow -E$, and $v^\mu \leftrightarrow {\tilde v}^\mu$).
However, we will have to take into account both degrees of freedom when computing physical observables.

In order to make contact with transport theory, one defines the (gauge-covariantly modified) Wigner transform of the the above two-point functions.
If $ X = \frac{1}{2} (x+y)$ and $s = x -y$ define the center of mass and relative coordinates, respectively, then
\be S_{E,v} (X,k) = \int d^4s e^{ik \cdot s} U \left(X,X+\frac{s}{2} \right) S_{E,v} (X + \frac{s}{2},X-\frac{s}{2}) U\left(X-\frac{s}{2},X \right) \ , \ee
where $U$ is the Wilson line,
\be U(x,y)=P \exp \left[ -i  e\int_\gamma dx^\mu A_\mu(x) \right] \ , \ee
and $P$ denotes path-ordering along the path $\gamma$ from $x$ to $y$. Using that 
\be
U(X,X+ \frac s2) U(X-\frac s2,X) \approx e^{i es \cdot A(X)} \ ,
\ee
then one can show that the introduction of the Wilson line allows  us to
define the Wigner function in terms of the kinetic momentum ${\bar k}^\mu = k^\mu -e A^\mu(X)$.
From now on, we will denote the kinetic momentum without the bar  to keep  the notation light.

We will focus on the construction of the transport equation associated with the vector and axial vector components of the the above two-point function, and define
\be
\label{the-current-standard}
{\rm Tr} ( \gamma^\mu S_{E,v} (X,k) ) = \sum_{\chi= \pm} {\rm Tr} ( \gamma^\mu P_\chi   \gamma_\nu   J_{E,v} ^{\nu,\chi} (X,k) )=  2 \sum_{\chi= \pm} J_{E,v} ^{\mu,\chi} (X,k) \ ,
\ee
where $\chi$ is an index that indicates the helicity/chirality  of the particle, and
\be
P_\chi = \frac{( 1 + \chi \gamma_5)}{2}
\ee
 is a chirality projector.

Now, simply by using that 
\be 
g^{\mu \nu} = P^{\mu \nu}_\perp + \frac{1}{2} \left( v^\mu {\tilde v}^\nu +  v^\nu {\tilde v}^\mu \right) \,,
\label{eq:momsplit}
\ee 
 one can decompose 
\be
\label{Wigner-current}
J^{\mu, \chi}_{E,v} (X,k) = v^\mu G^\chi_{E,v} (X,k) + \tilde{v}^\mu H^\chi_{E,v}(X,k)   + J^{\mu,\chi}_{(E,v), \perp}(X,k) \ .
\ee
Further, for the constraint $  \slashed{v} \chi_v = 0$ for particles,  one can deduce that $ H^\chi_{E,v}=0$. 
One can also show that $\langle \bar \chi_v(x) \gamma_\mu^\perp  \chi_v(x) \rangle =0$, and thus, $J^{\mu,\chi}_{(E,v), \perp}(X,k) =0$.

We will thus write our transport equations in terms of the two-point function
 \be 
\label{basic2GF}
 G_{E,v} (x,y) =  \langle \bar \chi_v (y)  \frac{\slashed{\tilde v}}{2}  {\chi}_v (x) \rangle \ee
and its (gauge-covariantly modified) Wigner transform.

 A basic ingredient to derive classical or semiclassical transport equations is to perform the gradient expansion,
 which assumes
\be \pa_X \ll \pa_s\ ,\ee
By doing this, we will consistently neglect gradients of the gauge fields. This does not mean that we are considering
only situations of  constant background fields, but rather  that their variation is consistently
neglected, as we will not take into account second order derivatives on $X$ of the two-point Green function.

\section{Derivation of the collisionless transport equation}
\label{Derivation}

\subsection{Computation using the OSEFT variables}
\label{OSEFTcomputation}

For our derivation, we substantially follow the approach of  Ref.~\cite{Son:2012zy},
where a chiral transport equation valid for Fermi systems at $T=0$ was derived from HDET \cite{Hong:1998tn}.
Actually, one of the motivations to develop OSEFT in Ref.\cite{Manuel:2014dza}  was to extend the validity of the same derivation at finite temperature,
where also antiparticles have to be taken into account. 
While in a system at finite density and vanishing temperature the Fermi sea provides a natural privileged frame, 
our derivation will be valid for an arbitrary frame.
 With some minor technical differences (the use of Dirac rather than Weyl fermions, use of local field redefinitions,
and consideration of nonhomogeneous distribution functions), we will find the final form of the chiral transport equation in an arbitrary frame,  respectful of reparametrization invariance, and therefore,
 Lorentz invariance. We will  point out an important difference from Ref.\cite{Son:2012zy} in our final results.

We start by considering the equations obeyed by the two-point Green's functions, as follows from the OSEFT Lagrangian.
To derive the collisionless transport equation it is enough to consider the  tree level equations. These can be expressed as 
\be \label{eq:eqs-1} 
\sum_{n=0} \left({\cal O}^{(n)}_x \right)\,S_{E,v} (x,y) =0
 \ , \ee
 and 
 \be \label{eq:eqs-2} 
\sum_{n=0}  S_{E,v} (x,y) ({\cal O}^{(n)}_y)^\dag  = 0
 \ , \ee
where from the OSEFT Lagrangian we can extract \footnote{Note that the difference in sign with respect to Ref.~\cite{Manuel:2016wqs} is due in the difference convention in the covariant derivative.}
\ba
{\cal O}_x^{(0)} &=& i\, v\cdot D \, \frac{\slashed{\tilde{v}}}{2} \ , \\
{\cal O}_x^{(1)} &=& - \frac {1}{2 E} \,\left( D_{\perp}^{2} + \frac e2 \sigma^{\mu \nu}_\perp F_{\mu \nu} \right) \frac{\slashed{\tilde{v}}}{2}\, \ , \\
{\cal O}_x^{(2)} &=&   -\frac{1}{4 E^{2}}  i \slashed{D}_\perp  (i {\tilde v} \cdot D) i \slashed{D}_\perp \frac{\slashed{\tilde{v}}}{2} 
\label{n=2Lag}
\\
\nonumber
&=& \frac{1}{8E^{2}}  \Big(\left[ \slashed{D}_{\perp}\,,\,\left[i\tilde{v}\cdot D\,,\,\slashed{D}_{\perp}\right]\right] +
\left\{ (  \slashed{D}_\perp)^2 , i\tilde{v}\cdot D \right\}  \Big)  \frac{\slashed{\tilde{v}}}{2}  
\,\ ,
\ea
and we limit our study to operators up to $1/E^2$ in the energy expansion.

It is  convenient to introduce local field redefinitions to eliminate the temporal derivative in Eq.~(\ref{n=2Lag}), as in Ref.~\cite{Manuel:2016wqs}, as these simplify quite a lot the 
 computations at higher orders~\footnote{Please note that in doing so one can ignore the Jacobian of the field redefinition in the computation of the generating functional of
 the Green's function~\cite{Arzt:1993gz} when considering on-shell Green's functions.}.   Local field redefinitions might not be respectful of RI if one considers off-shell quantities, but they will not affect the result
of on-shell quantities.
Thus, after the field redefinition
\begin{equation}
\label{LFR}
\chi_{v}\rightarrow\chi_{v}^{\prime}=\left(1 +\frac{   \slashed{D}_{\perp}^{2}}{8E^{2}}\right)\chi_{v}\ ,
\end{equation}
the second order differential operator becomes
\ba
{\cal O}_{x, \rm LFR}^{(2)} &=&  \frac{1}{8E^{2}}  \Big(\left[ \slashed{D}_{\perp}\,,\,\left[i\tilde{v}\cdot D\,,\,\slashed{D}_{\perp}\right]\right] -
\left\{  D_{\perp}^{2} + \frac e2 \sigma^{\mu \nu}_\perp F_{\mu \nu} ,\,\left(iv\cdot D-i\tilde{v}\cdot D\right)\right\}  \Big)  \frac{\slashed{\tilde{v}}}{2} \ .  \nonumber \\
\ea

We have checked that these two forms of the second-order Lagrangian lead to an equivalent form of the (on-shell) transport equation.

We now combine the sum and difference of Eqs.~(\ref{eq:eqs-1}) and  (\ref{eq:eqs-2}), and compute their Wigner transform. 
For every order in the energy expansion we define 
\be
\label{I-functions}
 I^{(n)}_{\pm} =  \int d^4s e^{i k\cdot s} \left({\cal O}^{(n)}_x  U(x,y) \, S_{E,v} (x,y) \pm S_{E,v} (x,y)U(x,y) {\cal O}^{(n) \dag}_y  \right) \ , \ee
however, note that these are  matrix equations in the Dirac subspace of the particles. 
In order to recover the transport equation we trace the above equations
\be
{\rm Tr} ( I^{(n)}_{\pm}) = \sum_{\chi=\pm} I^{(n)}_{\chi,\pm}  \ .
\ee

We can also derive separate equations for each helicity by multiplying by the appropriate chiral projector.

Furthermore, from Eqs.~(\ref{the-current-standard}) and (\ref{Wigner-current}) one can write
\be
G^\chi_{E,v}(X,k) = \frac 12 ( {\tilde v} \cdot J^\chi_{E,v})(X,k)  \ .
\ee

We leave for the Appendix \ref{compute-Is} some details of the computations, and present here our final results.
For $n=0$,
\begin{eqnarray}
I_{\chi,+}^{(0)}& =& 4 k \cdot v  \,  G^\chi_{E,v}(X,k) \ ,
 \\
 I_{\chi,-}^{(0)} &=& 2 i v_\mu [ \pa^\mu_X  -e  F^{\mu \nu} (X) \pa_{k, \nu} ] G^\chi_{E,v}(X,k)  \ ,
\end{eqnarray}
for $n=1$, 
\begin{eqnarray}
 I_{\chi,+}^{(1)} & = & \frac{2}{E}  \left(  k^2_{\perp} - \frac{e \chi}{4} \epsilon^{\alpha \beta \mu \nu} {\tilde v}_\beta v_{\alpha} F^\perp_{\mu \nu}
  \right) G^\chi_{E,v}(X,k) \ ,
  \\
 I_{\chi,-}^{(1)} & = & 2 \frac{i}{E}  {k}_{\perp}^{\mu} \  [\pa_{X,\mu}-  e F_{\mu \nu} \pa_k^\nu] \ G^\chi_{E,v}(X,k) \ ,
\end{eqnarray}
while for $n=2$, one gets
\be
I_{\chi,+}^{(2)} = 
-\frac{2}{E^2} \left( \left[  k^2_{\perp}  - \frac{e \chi}{4} \epsilon^{\alpha \beta \mu_\perp \nu_\perp} {\tilde v}_\beta v_{\alpha} F_{\mu \nu} \right]  \frac{{\tilde v}\cdot k - v \cdot k}{2}  + \frac{e \chi}{4}     \epsilon^{\alpha \beta \mu_\perp \nu_\perp} {\tilde v}_\beta v_{\alpha} F_{\nu \rho} {\tilde v}^\rho k_{\mu}   \right)  G^\chi_{E,v}(X,k) \ ,
\label{nightmare+}
\ee
and 
\ba
I_{\chi,-}^{(2)} &= &
 \frac{2}{E^2} \left( - k^\mu_{\perp} \frac{{\tilde v}\cdot k - v \cdot k}{2}   + \frac 14 \left[  k^2_{\perp}  - \frac {e\chi}{4}  \epsilon^{\alpha \beta \delta \gamma} {\tilde v}_\beta v_{\alpha} F^\perp_{\delta \gamma} \right] (v^\mu - {\tilde v}^\mu )
- \frac{e \chi}{8} \epsilon^{\alpha \beta \mu_\perp \nu_\perp} {\tilde v}_\beta v_{\alpha} F_{\nu \rho} {\tilde v}^\rho  \right) 
\nonumber
\\
& \times& i [ \pa_{X,\mu}  -e  F^{\sigma}_\mu (X) \pa_{k, \sigma} ]  G^\chi_{E,v}(X,k) \ .
\label{nightmare-}
\ea

We can check that, when computed in the static frame defined by fixing the frame vector as $u^\mu=(1,0,0,0)$, and using Eq.~(\ref{frame-vel}), our results agree with those computed from HDET  in Ref.~\cite{Son:2012zy}
if we replace the chemical potential $\mu$ by the energy $E$, except in what follows.
With the local field redefinition, the factor multiplying the time derivative in the transport equation is 1, while without it one gets a nontrivial factor. We have checked that the same
equation is obtained if we normalize the transport equation of Ref.~\cite{Son:2012zy} so as to obtain the same normalization of the time derivative term.
We, however, disagree in the numerical factor of  the piece  proportional to $F_{\nu \rho} {\tilde v}^\rho$ in  Eqs.~(\ref{nightmare+}) and (\ref{nightmare-}), in what it is apparently an algebraic mistake.
The numerical factors found above turn out to be essential to deriving both the proper form of the dispersion relation, and the consistent form of the  anomaly equation.

\subsection{Going backward to the original variables}
\label{backwards} 

Having derived the relevant equations in terms of the OSEFT variables, let us now go back and express them in terms of the original momenta of the full theory.

\subsubsection{Dispersion relation}

The dispersion relation is fixed after imposing

\be 
\label{on-shell-conditon} 
I_{\chi,+}^{(0)} + I_{\chi,+}^{(1)} + I_{\chi,+}^{(2)}=0 \ ,
\ee
which suggests that the Wigner function can be written as
\be
\label{Wigner-function}
 G^\chi_{E,v}(X,k)= 2 \pi \delta ( K^\chi) f_{E,v}^\chi (X,  k) \ ,
\ee 
where $f_{E,v}^\chi (X,  k)$ is the particle distribution function, and  we have introduced a $(2 \pi)$ factor in order to reproduce, to leading order, the expected density in a QED plasma. We keep the labels 
$E$ and $v$ in the distribution function, as this function will depend on
the on-shell variables; see for example Ref.\cite{Manuel:2016wqs}, where it was explicitly seen that close to equilibrium the on-shell energy acts as a sort of chemical potential
for the residual momentum. The function $K^\chi$ fixes then the dispersion relation, to the order considered, and can be
read from the $I_{\chi,+}$ functions. In particular, up to order $n=2$,
\ba
 K^\chi &= & 2 k \cdot v + \frac{1}{E}  \left(  k^2_{\perp} - \frac{e \chi}{4} \epsilon^{\alpha \beta \mu \nu}  {\tilde v}_\beta v_{\alpha} F^\perp_{\mu \nu}
  \right) \nonumber \\
  & - &\frac{1}{E^2} \left( \left[  k^2_{\perp}  - \frac{e \chi}{4} \epsilon^{\alpha \beta \mu_\perp \nu_\perp}  {\tilde v}_\beta v_{\alpha} F_{\mu \nu} \right]  \frac{{\tilde v}\cdot k - v \cdot k}{2}  + \frac{e \chi}{4}     \epsilon^{\alpha \beta \mu_\perp \nu_\perp} 
   {\tilde v}_\beta v_{\alpha} F_{\nu \rho} {\tilde v}^\rho k_{\mu}   \right)  \ .
\ea
Note that we could replace $\epsilon^{\alpha \beta \mu \nu} {\tilde v}_\beta v_\alpha = 2 \epsilon^{\alpha \beta \mu \nu} u_\beta v_\alpha$ in the above expression.
The on-shell constraint can be solved to different orders in the energy expansion. To leading order it is simply
\be 
\label{ondell0}
2 k \cdot v  = 0 \ ,
\ee
while at the following order,
\be
\label{on-shell1}
2 k \cdot v + \frac{1}{E}  \left(  k^2_{\perp} - \frac{e \chi}{4} \epsilon^{\alpha \beta \mu \nu} {\tilde v}_\beta v_{\alpha} F^\perp_{\mu \nu} \right) = 0 \ ,
\ee
 showing that $(v\cdot k)$ turns out to be subleading in the $1/E$ expansion when taken on shell.

It turns out convenient to express the on-shell constraint in terms of the original momentum $q^\mu$. Then one can check that it leads to the constraint
\be
\label{eq:onshell}
q^2 - e S_\chi^{\mu \nu} F_{\mu\nu} = 0  \, ,
\ee
where $S_\chi^{\mu \nu} $ is the spin tensor defined as 
\be
S_\chi^{\mu \nu} =  \chi\frac{ \epsilon^{\alpha \beta \mu \nu} u_\beta q_\alpha}{2 (q \cdot u)}  \ ,
\ee
if solved up to order $1/E^2$ in the OSEFT variables. To see this, we can express \Eq{eq:onshell} in terms of on-shell and residual momenta. 
Using
\be
E_q \equiv  q \cdot u = E+  k \cdot u   \ ,
\ee
and also that we can write  for the residual momentum  
\be
k^\mu = k^\mu_\perp + \frac{1}{2} (v\cdot k)\tilde{v}^\mu + \frac{1}{2} (\tilde{v}\cdot k)v^\mu \,, \qquad k^2 = k_\perp^2 + (v\cdot k) (\tilde v \cdot k) \,,
\ee 
then the spin tensor can be written as
\be
S_\chi^{\mu\nu} =  \frac{\chi}{2} \epsilon^{\alpha \beta \mu \nu} u_\beta
\left( v_\alpha + \frac{k^\perp_\alpha}{E} \right)  + {\cal O}\left(\frac{1}{E^2}\right)  \,.
\ee

We can then easily obtain
\be
q^2 - e S_\chi^{\mu \nu} F_{\mu\nu} =  2E \left[ v \cdot k + \frac{1}{2E} \left( k^2_\perp -e S_\chi^{\mu\nu} F_{\mu\nu} \right) \left( 1 - \frac{ ( {\tilde v} \cdot k)}{2E} \right) \right]  + {\cal O}\left(\frac{1}{E^2}\right) \,,
\label{eq:drOSE}
\ee
where in the last expression we used Eq.~(\ref{on-shell1}) and the fact that we are considering expansions in powers of $1/E$. 
Furthermore, employing once again the decomposition in Eq.~(\ref{eq:momsplit}) both for $k_\alpha$ and $F_{\mu\nu}$, we can express $S_\chi^{\mu \nu} F_{\mu\nu}$ in terms of the OSEFT variables
\be
S_\chi^{\mu\nu} F_{\mu\nu} = \frac{\chi}{2} \epsilon^{\alpha \beta \mu \nu} u_\beta \left( v_\alpha + \frac{k^\perp_\alpha}{E} \right) 
\left( F_{\mu\nu}^\perp + F_{\mu_\perp \rho}{\tilde v}^\rho v_\nu + F_{\mu_\perp \rho}v^\rho{\tilde v}_\nu \right) +  {\cal O}\left(\frac{1}{E^2}\right)
\label{eq:SF}
\ee
Finally, we can replace the above vector $u_\beta$  by $\tilde{v}_\beta /2$, the difference being a higher $1/E$ effect. This can be checked by noticing that $v^\mu A_\mu \ll {\tilde v}^\mu A_\mu$. Note that the condition Eq.~(\ref{ondell0}) involves the kinetic, rather than canonical, momentum, which implies that not all the vector gauge field components are equally relevant in the $1/E$ expansion.

Under these conditions one can then check that \Eq{eq:drOSE} becomes exactly $ E K^\chi $.   \Eq{on-shell-conditon} thus enforces the on-shell condition \Eq{eq:onshell}, as anticipated.

Thus, in returning to the original variables, we will identify, to order $n=2$ accuracy in the $1/E$ expansion,
\be
G^\chi_{E,v}(X,k)= (2 \pi)  \delta ( K^\chi) f_{E,v}^\chi (X,  k) =  (2 \pi) E \,\delta ( E K^\chi) f_{E,v}^\chi (X,  k) =
\pi  E\,  \delta_+(Q^\chi) f^\chi (X, q) \ ,
\ee
where we have defined
\be
\delta_+(Q^\chi)  = \delta \left(  q^2 -  e S^{\mu \nu}_\chi F_{\mu \nu}  \right) 2 \theta(E_q ) \ .
\ee

When the Wigner function is expressed in terms of the original variables, there is still an $E$ dependence.
In explicit computations of physical parameters, such as the vector current (see Sec.~\ref{Consistentcurrent+anomaly}),  this $E $ dependence  disappears when one
finally expresses the whole current in terms of the original variables.

\subsubsection{Transport equation}

The transport equation is obtained from

\be 
\label{transport-eq} 
I_{\chi,-}^{(0)} + I_{\chi,-}^{(1)} + I_{\chi,-}^{(2)}=0  \ .
\ee

We will express the transport equation  in terms of the original momentum $q^\mu$.
Let us define the vector
\be
\label{originalvel}
v_\mu^q \equiv \frac{q^\mu}{E_q}  = \frac{E}{E_q} v^\mu + \frac{k^\mu}{E_q} \ ,
\ee
which satisfies $u \cdot v_q = 1$.
 In the absence of gauge fields this vector can be written as
\be
v_\mu^q = v^\mu + \frac{k^\mu - v^\mu (k \cdot u)}{E} -  (k \cdot u) \frac{k^\mu - v^\mu (k \cdot u)}{E^2} + \cdots
\ee
If we further consider the on-shell condition at lowest order $v \cdot k =0$, then
\be
k^\mu - v^\mu (k \cdot u) \Big |_{\rm o.s.} = k^\mu_\perp \ ,
\ee
and it is not difficult to realize that
\be
\label{os-nogauge}
v_\mu^q \Big |_{\rm o.s.} = v^\mu +  \frac{k^\mu_\perp}{E} - (k \cdot {\tilde v}) \frac{k^\mu_\perp}{2E^2} +\frac{v^\mu -{\tilde v}^\mu}{4 E^2} k^2_\perp + {\cal O}(\frac{1}{E^3}) \ .
\ee

If we now we include the gauge fields, after using Eq.~(\ref{on-shell1}) we then get 
\be
\label{onshellvector}
v_\mu^q \Big |_{\rm o.s.} = v^\mu +  \frac{k^\mu_\perp}{E} - (k \cdot {\tilde v}) \frac{k^\mu_\perp}{2E^2} +\frac{v^\mu -{\tilde v}^\mu}{4 E^2} \left(k^2_\perp -
\frac{e \chi}{4} \epsilon^{\alpha \beta \mu \nu} {\tilde v}_\beta v_{\alpha} F^\perp_{\mu \nu} \right)
 + {\cal O}(\frac{1}{E^3}) \ 
\ee
which is the combination that appears in the $I_{\chi,-}$ functions.

If we  define 
\be
\Delta^\mu \equiv \pa^\mu_X  -e  F^{\mu \nu} (X) \pa_{q, \nu} \,,
\ee
one can write the transport equation in terms of the original variables as
\be
\label{CCTEq-1}
\left( v_\mu^q - \frac{e}{2 E^2_q}S_\chi^{\mu \nu}F_{\nu \rho} \left(2 u^\rho - v^\rho_q \right) \right)
\Delta_\mu f(X,q) \delta_+ (Q)
 = 0  \ ,
 \ee
where we have used that ${\tilde v}^\rho = 2 u^\rho - v^\rho_q$ in the last term only.  In the absence of the $1/E_q$ corrections, \Eq{CCTEq-1} corresponds to
a classical transport equation of a charged fermion in the collisionless limit \footnote{In a previous version of this paper, we used the vector (\ref{os-nogauge}) in the final
form of the equation. The equation gets a much more compact form if it is expressed in terms of Eq.~(\ref{onshellvector}).}.

After taking into account the on-shell condition, Eq.~(\ref{CCTEq-1}) is similar, but not identical, to the one proposed in Ref.~\cite{Hidaka:2017auj}, see also
Refs.~\cite{Hidaka:2016yjf,Hidaka:2018ekt}, if we identify their frame vector $n^\mu$ with our $u^\mu$. For homogeneous backgrounds, 
Eq.~(\ref{CCTEq-1}) contains a term, the  piece proportional to $S_\chi^{\mu \nu}F_{\nu \rho}  v^\rho_q$, which is absent in Eq.~(11) of Ref.~\cite{Hidaka:2017auj}. It could be eliminated by introducing a new term in the 
OSEFT Lagrangian at order $1/E^2$ , namely, the same that appears in  Eq.~(\ref{n=2Lag}), but changing the $({\tilde v} \cdot D)$ by $(v \cdot D)$. However, this could only be done at the expense of breaking reparametrization invariance and, ultimately, Lorentz invariance.

For nonhomogeneous backgrounds, Eq.~(11) of Ref.~\cite{Hidaka:2017auj} 
kept some gradient terms of the gauge fields and frame vector. The gradient expansion used to reach to the above transport equation was made by neglecting
gradients of the electromagnetic fields (see Appendix ~\ref{compute-Is}), which would otherwise naturally emerge in the computations of the functions $I_{\chi, -}$; thus, not all the
gradient terms were kept in Refs.~\cite{Hidaka:2016yjf,Hidaka:2018ekt}, and in a close to thermal equilibrium situation, it might be nonconsistent to keep those gradient terms
while neglecting $\pa^2_X G$.

Let us consider now our covariant relativistic equation and write it in the frame $u^\mu =(1,0,0,0)$. In this frame, $F^{i0}= E^i$, $F^{ij} = - \epsilon^{ijk} B^k$, and also
\be
S_\chi^{\mu \nu} \rightarrow S^{ij}_\chi = \chi \frac{ \epsilon^{ijk}q^k}{2 q_0} \ , \qquad  S_\chi^{\mu \nu} F_{\mu\nu} = - \chi \, {\bf B} \cdot \frac{\bf q}{q_0} \ .
\ee

After considering the on-shell condition, it is not difficult to arrive at
\be
\label{staticCKT}
\left ( \Delta_0 + {\bf  \hat q}^i \left (1 + e \chi \frac{ \bf B \cdot \bf \hat q}{2 q^2} \right) \Delta_i  + e \chi \frac{\epsilon^{ijk} E^j \hat{q}^k - B^i_{\perp,\bf q}}{4 q^2}  \Delta_i
\right) f^\chi(X, {\bf q}) =0   \ ,
\ee
where we have defined $B^i_{\perp,\bf q} \equiv B^i -{\bf \hat q}^i ({\bf B} \cdot {\bf \hat q})$. This equation differs from Eq.~(13) of Ref.~\cite{Hidaka:2016yjf}, which for homogeneous backgrounds reads
\be
\left ( \Delta_0 + {\bf  \hat q}^i \left (1 + e \chi \frac{ \bf B \cdot \bf \hat q}{2 q^2} \right) \Delta_i  + e \chi \frac{\epsilon^{ijk} E^j \hat{q}^k }{2 q^2}  \Delta_i
\right) f^\chi(X, {\bf q}) =0   \ .
\ee

Eq.~(\ref{staticCKT}) also differs from the transport equation described in Sec.~IIB of Ref.~\cite{Son:2012zy}, although that equation leads to the covariant chiral
anomaly equation, while ours leads to the consistent form of the chiral anomaly equation, as we discuss in the following section.

\section{Consistent current and  chiral anomaly equation}
\label{Consistentcurrent+anomaly}

In this section, we compute both the consistent electromagnetic and chiral currents.
For the computation of the latter, the best option is to  introduce an artificial chiral gauge field $A^5_{\mu}$ and an artificial gauge field tensor $F_{\mu \nu}^5$, which are finally sent to zero, 
as advocated in Ref.~\cite{Landsteiner:2016led}, and in Ref.~\cite{Gorbar:2016ygi}, for example.
Thus we assume that the original QED Lagrangian reads
\be
{\cal L} =  \sum_{E, v} \left ( \bar \psi_{v, \tilde v} (x) \, i \gamma^\mu \left( \pa_\mu + i e A_\mu + i e \gamma_5 A_\mu^5\right) \psi_{v, \tilde v} (x) \right ) \,.
\ee
One can proceed with the same derivation of the OSEFT Lagrangian in the presence of the chiral field. After introducing the chiral projectors,
it is not difficult to realize that all our equations remain valid if we replace
\be
A_\mu  \rightarrow A_\mu + \chi A_\mu^5 \ ,  \qquad   F_{\mu \nu} \rightarrow F_{\mu \nu} + \chi F^5_{\mu \nu} \ ,
\ee
in all our final formulas, in agreement with the prescription of Ref.~\cite{Gorbar:2016ygi}. 

The electromagnetic and chiral currents are obtained from the OSEFT action, simply by performing the functional derivatives

\begin{equation}
j^\mu(x) = - \frac{ \delta {\cal S}}{\delta A_\mu (x)} \ , \qquad  j^5_\mu(x) = - \frac{ \delta {\cal S}}{\delta A^5_\mu (x)} \ ,
\end{equation}
respectively. Alternatively, one could start with the QED currents, and plug the explicit expression of the Dirac fields in Eq.~(\ref{eq:Fields})
to finally write the current in terms of the OSEFT fields.
For example, considering only the contribution of the particles
\be
{\bar \psi}_{ v, \tilde v}(x) \gamma^\mu\psi_{ v, \tilde v} (x) \rightarrow \left( \bar \chi_{v}(x) +  \bar H_{\tilde v}^{(1)}(x)\right) \gamma^\mu \left(  \chi_{v}(x) +  H_{\tilde v}^{(1)}(x)\right) \equiv  j^\mu (x)
\ee
Using the expression of the $H_{\tilde v}^{(1)}$ of Ref.~\cite{Manuel:2014dza} generalized to an arbitrary frame, we find 
\ba
j^\mu (x) &= &
v^\mu \bar \chi_{v} \frac{\slashed{\tilde v}}{2} \chi_v
+
\frac{1}{2E}\left( \bar \chi_{v} \gamma^\mu_\perp i \slashed{D}_{\perp}\frac{\slashed{\tilde v}}{2} \chi_v +  \bar \chi_{v}  (i  \overleftarrow{\slashed{D}})_{\perp} \gamma^\mu_\perp \frac{\slashed{\tilde v}}{2} \chi_v \right) 
\nonumber
 \\
&-& \frac{{\tilde v}^\mu}{4 E^2} \left( \bar \chi_{v}  (i  \overleftarrow{\slashed{D}})_{\perp} (i  \slashed{D})_{\perp}  \frac{\slashed{\tilde v}}{2} \chi_v \right)
+ \frac{{ v}^\mu}{8 E^2} \left( \bar \chi_{v} ( \overleftarrow{\slashed{D}})^2_{\perp}  \frac{\slashed{\tilde v}}{2} \chi_v   + \bar \chi_{v}  ( \slashed{D})^2_{\perp}  \frac{\slashed{\tilde v}}{2} \chi_v  \right) \nonumber \\
&-& \frac{1}{4 E^2} \left( \bar \chi_{v}  (i {\tilde v}\cdot D) \gamma^\mu_\perp (i  \slashed{D})_{\perp}  \frac{\slashed{\tilde v}}{2} \chi_v  +
\bar \chi_{v}   (i  \overleftarrow{\slashed{D}})_{\perp} (i {\tilde v}\cdot \overleftarrow{D}) \gamma^\mu_\perp \frac{\slashed{\tilde v}}{2} \chi_v  \right) + {\cal O}(\frac {1}{E^3})
 \ ,
\ea
where we have to take into account the local field redefinition, Eq.~(\ref{LFR}), so as to compute the current in the same way as the corrections to the
transport equations. A completely analogous computation can be carried out for the chiral current.

At leading order in the energy expansion, one can immediately express the current in terms of the two-point function.
After a Wigner transform, one finds
\be
j^\mu_{(0)} (X) = e  \sum_{E, v, \chi}  \int \frac{d^4 k}{(2 \pi)^4 } v^\mu \,  2 G_{E,v}^\chi(X,k) \ .
\ee
We can use now the explicit form of the Wigner function at order $n=0$; see Eq.~(\ref{Wigner-function}). If we further make the identification \cite{Luke:1999kz,Manuel:2016wqs}
\be
\sum_{E, v}  \int \frac{d^4  k}{(2 \pi)^4 } \equiv \int \frac{d^4  q}{(2 \pi)^4 }  \,,
\ee
then, at leading order, the current is expressed as

\be
j^\mu_{(0)} (X) = e  \sum_{\chi = \pm} \int \frac{d^4  q}{(2 \pi)^3 } \, 2 \theta({E_q}) \,\delta(q^2)  q^\mu  f^\chi(X, q) \ ,
\ee
where  we have approximated $ E v^\mu \approx q^\mu$ at leading order, and it is understood that the on-shell condition is taken to leading order, thus, without the gauge field
contribution.
Similarly, the axial current at leading order reads
\be
j^\mu_{5,(0)} (X) = e  \sum_{\chi =\pm } \chi \int \frac{d^4  q}{(2 \pi)^3 } \, 2 \theta({E_q}) \,\delta(q^2)  q^\mu  f^\chi(X, q) \ .
\ee

At the following orders in the energy expansion, and due to the presence of derivative terms in the explicit expression of the current, a point-splitting regularization is needed.
This means that we take the field $\bar \chi_v$ at the value $y$. We then perform the (gauge-covariantly modified) Wigner transform, together with the derivative expansion, to finally
take the limit $y \rightarrow x$. Note that this point-splitting regularization is only needed to properly define the Wigner transform (see, for example, the scalar QED example explained in Ref.\cite{Blaizot:1999xk}
for the proper definition of the current) and not to regulate ultraviolet problems, which are absent in the two-point function we are studying.

If one considers corrections up to order $n=2$, then the vector current reads

\ba
\label{OSEFT-curren2}
j_{(2)}^\mu(X) &=& e  \sum_{E, v, \chi}  \int \frac{d^4 k}{(2 \pi)^4 } \Bigg \{ \left(v^\mu +  \frac{k^\mu_\perp}{E} - (k \cdot {\tilde v}) \frac{k^\mu_\perp}{2E^2} +\frac{v^\mu -{\tilde v}^\mu}{4 E^2} \left(k^2_\perp - \frac{e \chi}{4} \epsilon^{\alpha \beta \mu \nu} {\tilde v}_\beta v_{\alpha} F^\perp_{\mu \nu} \right)\right) \Bigg.
\nonumber \\
&-& \frac{\chi}{4E}\left( \epsilon^{\mu \nu \alpha \beta} {\tilde v}_\alpha v_\beta    - \frac{(k \cdot {\tilde v})}{2 E} \epsilon^{\mu \nu \alpha \beta} {\tilde v}_\alpha v_\beta\right)   [ \pa_\nu^X  -e  F_{ \nu \sigma}  \pa_{k}^\sigma ]  + \frac{\chi}{8E^2} \epsilon^{\mu \nu \alpha \beta} {\tilde v}_\alpha v_\beta  k_\nu
{\tilde v}^\rho [ \pa_\rho^X  -e  F_{ \rho \sigma}  \pa_{k}^\sigma ]  
\nonumber
\\
&+&   \Bigg.\frac{e \chi}{ 8 E^2} \epsilon^{\mu \rho  \alpha \beta}{\tilde v}_\alpha v_\beta F_{\rho \sigma} {\tilde v}^\sigma  \Bigg \} 2 G_{E,v}^\chi(X,k) \ ,
\ea
which, if converted to the original momentum, reads 
\be
\label{current}
j_{(2)}^\mu(X) = e \sum_{ \chi= \pm} \int \frac{d^4 q}{(2 \pi)^3 }  \ \Bigg \{ q^\mu + S_\chi^{\mu \nu} \Delta_\nu 
-
\frac{e}{2 E_q} S^{\mu \nu}_\chi F_{\nu \rho} (2 u^\rho - v^\rho_q) \Bigg \} f^\chi(X,q) \,\delta_+(Q^\chi)  \ .
\ee
For the axial current we get the same expression but the whole integral is multiplied by $\chi$.

In order to get the complete current, the antiparticle contribution has to be added. As mentioned in Sec.~\ref{WignerSection} this can be recovered from the OSEFT particle contribution, Eq.~(\ref{OSEFT-curren2}), by simply replacing
$v^\mu \leftrightarrow {\tilde v}^\mu$ and $E \rightarrow -E$.

Let us consider the current associated with one single value of the chirality. Using the transport equation (\ref{CCTEq-1}) and the antisymmetry of the
spin tensor, it is not difficult to deduce
\ba
\label{conse-current}
\pa_\mu j_\chi^\mu(X) &=& e^2   \int \frac{d^4 q}{(2 \pi)^3 } \Bigg \{ q^\mu + S_\chi^{\mu \nu} \Delta_\nu -
\frac{e}{2 E_q} S^{\mu \nu}_\chi F_{\nu \rho} (2 u^\rho - v^\rho_q) \Bigg \} F_{\mu \lambda}\frac{\pa}{\pa q^\lambda}( f^\chi \, \delta_+(Q^\chi) ) \ .
\ea

To deduce the form of the chiral anomaly, we will now consider the frame $u^\mu =(1,0,0,0)$, as then the analysis simplifies quite a lot.
We will also consider the situation where, to leading order, the distribution function corresponds to a thermal distribution function, with a chemical potential
that depends on the chirality: that is, there is a  fermion chiral imbalance in the system. The proof, however, can also be extended to distribution functions which, when the on-shell
condition to leading order is considered, are parity invariant.
 One can express the integral on the rhs. of Eq.~(\ref{conse-current}), after taking into account the on-shell condition,
as a surface integral. As the distribution function vanishes for $|{\bf  q}| \rightarrow \infty$, the only nonvanishing contribution arises for low values of the momenta,
where the quasiparticle picture breaks down. We proceed as in Ref.~\cite{Stephanov:2012ki}, and Refs.~\cite{Manuel:2013zaa,Manuel:2014dza},
 and define a sphere centered in $|{\bf  q}| =0$ of radius $R$
and then compute the only nonvanishing surface integral
\ba
\pa_\mu j_\chi^\mu(X) &=& - e^2 \chi \lim_{R \rightarrow 0}\left( \int \frac{d{\bf S}_R}{(2\pi)^3} \cdot {\bf E} \, \frac{ {\bf \hat q} \cdot {\bf B}}{4 R^2}  f^\chi(|{\bf  q}| =R)  -
\int \frac{d{\bf S}_R}{(2\pi)^3} \cdot \, \frac{ {\bf \hat q}}{4R^2}  {\bf E}\cdot {\bf B} f^\chi(|{\bf  q}| =R)  \right)
\nonumber
 \\
&=&  e^2 \chi\frac{ {\bf E}\cdot {\bf B} }{2 \pi^2} \frac 16  f^\chi(|{\bf  q}| =0) \ .
  \ea

At this point, we should consider the contribution of all the chiralities, of both fermions and antifermions  so as to obtain the full complete
contribution to the axial and vector currents. We thus assume the following fermion and antifermion distribution functions,
\be
 f^\chi(|{\bf  q}|) =\frac{1}{e^{({|\bf  q}| -\mu_\chi)/T} +1} \ , \qquad   {\tilde f}^\chi(|{\bf  q}|) =\frac{1}{e^{({|\bf  q}| +\mu_\chi)/T} +1}  \ ,
\ee
respectively,
to obtain the nonconservation of the chiral current
\be
\label{chiral-consistent-ano}
\pa_\mu {\cal J}^\mu _5(X) =   \frac 13 \frac{e^2}{2 \pi^2} \left( {\bf E}\cdot {\bf B}  + {\bf E}_5\cdot {\bf B}_5 \right) \ .
  \ee
The vector current also has a quantum anomaly also in the presence of chiral gauge fields
\be
\label{vector-consistent-ano}
\pa_\mu {\cal J}^\mu(X) =   \frac 13 \frac{e^2}{2 \pi^2} \left( {\bf E_5}\cdot {\bf B}  + {\bf E} \cdot {\bf B}_5 \right) \ .
  \ee

Eq.~(\ref{chiral-consistent-ano}) gives account of the consistent form of the chiral anomaly equation, rather than its covariant form.
We refer the reader to the excellent review \cite{Landsteiner:2016led} that gives very clear explanations about these two different forms of the quantum anomaly.
After defining our currents as functional derivatives of the action, one cannot get anything else than the consistent currents.
Unfortunately, the vector current  is also nonconserved.
It is possible to add the so-called Bardeen counterterms \cite{Bardeen:1984pm} to the quantum action
\be
e^2 \int d^4 x \, \epsilon^{\mu \nu \rho \lambda} A_\mu A_\nu^5 \left( c_1 F_{\rho \lambda} + c_2  F^5_{\rho \lambda} \right) \ ,
\ee
with the choice $c_1=\frac{1}{12 \pi^2}$ and $c_2=0$,  and then one can get a vector conserved current  \cite{Landsteiner:2016led}.

Previous approaches to CKT have shown to provide both the covariant currents and also the covariant form of the chiral anomaly~\cite{Manuel:2014dza,Manuel:2016wqs,Manuel:2016cit}; see also  Ref.~\cite{Gorbar:2016ygi}.
One can relate the consistent and covariant currents by adding Chern-Simons currents  \cite{Landsteiner:2016led}.

\section{Side jumps derived from reparametrization invariance of the OSEFT}
\label{Sidejumps=RI}

Once we know how the fields of the OSEFT behave under the three types of RI transformations, we can deduce how the different two-point functions
behave under the same transformations. Then, after performing the (gauge-covariantly modified) Wigner transform and a gradient expansion, we can
deduce how the distribution function behaves under the same sort of transformations.

It is actually easy to show that under the  type I and type III  symmetries of RI the distribution function in the OSEFT remains invariant.
For example, under type I symmetry the  basic two-point function transforms as (see Table~\ref{RI-transfor})
\be 
 \langle  \bar \chi_v (y) \frac{\slashed{\tilde v}}{2}  \chi_v (x) \rangle ' \rightarrow   
  \langle \bar \chi_v (y) \left(1+ \frac 14 \slashed{\tilde v} \slashed{\lambda}_{\perp}  \right)  \frac{\slashed{\tilde v}}{2} \left( 1 + \frac 14  \slashed{\lambda}_{\perp} \slashed{\tilde v} \right)\chi_v(x) \rangle =
   \langle  \bar \chi_v (y) \frac{\slashed{\tilde v}}{2}  \chi_v (x) \rangle 
   \ , \ee
where we have used that $\slashed{\lambda}_\perp \slashed{\tilde v}= - \slashed{\tilde v} \slashed{\lambda}_\perp$, and $\slashed{\tilde v} \slashed{\tilde v} =0$.
It then follows that 
\be
( f^\chi_{E,v}(X,  k) )' = f^\chi_{E,v} (X, k) 
\ . \ee
under a type I transformation.
Similarly, it is possible to show that the distribution function does not change under a type III transformation.

The Green function (\ref{basic2GF}) used in our derivation of the transport equation has, however, a nontrivial transformation under type II symmetry. Using
the  transformation rules of Table~\ref{RI-transfor}, we obtain

\ba 
\langle  \bar \chi_v (y)  \frac{\slashed{\tilde v}}{2} \chi_v (x) \rangle '  & \rightarrow &  \langle \bar \chi_v (y)  \frac{\slashed{\tilde v} +\slashed{\epsilon}_\perp}{2} \chi_v(x) \rangle 
\\
\nonumber 
&+& 
 \langle \bar \chi_v (y) \left( \frac{(i \overleftarrow{\slashed{D}}_{\perp,y} )^\dag \slashed{\epsilon}_\perp^\dag }{2E} \right)      \frac{\slashed{\tilde v}}{2} \chi_ v(x) \rangle +
\langle \bar \chi_v (y) \frac{\slashed{\tilde v}}{2}  \left( \frac 12 \frac{ \slashed{\epsilon}_\perp   i \slashed{D}_{\perp,x}}{2E} \right)\chi_ v(x) \rangle + {\cal O} (\frac{1}{E^2}) \ . \ea

In OSEFT $\langle \bar \chi_v (y) \gamma^\mu_\perp \chi_v(x) \rangle =0$. 
After the Wigner transform, together with the gradient expansion, we end up with
\be 
(G^\chi_{E,v}(X, k))'  \rightarrow G^\chi_{E,v} (X,k) - \frac{1}{2E} k_\perp \cdot \epsilon_\perp G^\chi_{E,v} (X,k) - \frac{\chi}{E} \epsilon^{\mu_\perp \nu_\perp \alpha \beta} 
v_\alpha {\tilde v}_\beta \epsilon^\perp_\nu (\pa^X_\mu  -e F_{\mu \lambda} \pa_k^\lambda ) G^\chi_{E,v} (X,k) 
 \ , \ee

Taking into account the definition of the two-point function at order $1/E$ involves the current density that might be computed [see the integrand of  Eq.~(\ref{OSEFT-curren2})  at order $1/E$] as
\be
G^\chi_{E,v}(X, k) = \frac 12   {\tilde v}_\mu \cdot ( v^\mu + \frac{k^\mu_\perp}{E} + \cdots) (2 \pi ) 
f^\chi_{E,v}(X,k) \delta_+(K^\chi) \ ;
\ee
this implies that the distribution function should  change as
\be
(f^\chi_{E,v}(X, k))'  \rightarrow f ^\chi_{E,v} (X,k) - \frac{\chi}{E} \epsilon^{\mu_\perp \nu_\perp \alpha \beta} 
v_\alpha {\tilde v}_\beta \epsilon^\perp_\nu (\pa^X_\mu  -e F_{\mu \lambda} \pa_k^\lambda ) f^\chi_{E,v} (X,k) \ ,
\ee
under a type II transformation.

  In terms of the original variables, one then gets
\be
\label{side-jumpf}
(f^\chi(X, q))'  \rightarrow f ^\chi (X,q) - \frac{1}{E_q}  S^{\mu \nu}_\chi \epsilon_\nu^\perp \Delta_\mu 
f^\chi (X,q) + {\cal O} \left(\epsilon_\perp^2, \frac{1}{E_q^2}\right) \ .
\ee
Taking into account that $\epsilon^\mu_\perp/2 = u'^\mu - u^\mu$, we see that Eq.~(\ref{side-jumpf}) agrees with the infinitesimal form of the side-jump transformation
first discussed in Ref.~\cite{Chen:2015gta} in the absence of gauge fields, later generalized in the presence of the gauge fields in Ref.\cite{Hidaka:2016yjf}.

\section{Discussion}
\label{discussion}

We have derived  from OSEFT the corrections to the classical transport equations  associated with on-shell massless charged fermions and antifermions. We have 
seen how from the proposed equations one can derive the consistent form of the chiral anomaly equation when considering a chiral imbalance system in thermal
equilibrium.
Our formulation turns out to be the proper generalization of the HDET approach to chiral transport theory of Ref.~\cite{Son:2012zy}, but valid also for finite temperature
systems and formulated in an arbitrary frame. The study of reparametrization invariance of the theory allows
us to claim that the results are consistent with Lorentz symmetry, even if the kinetic equation depends on a frame vector. We have also deduced the side-jumps
of the distribution function of the theory from the transformation rule under RI of the OSEFT quantum fields.

Let us insist that when we consider the frame vector as $u^\mu =(1, {\bf 0})$, our equations almost agree with those of  Ref.~\cite{Son:2012zy}, except in a couple
of factors, in what apparently was an algebraic mistake.  It is, however, important to stress that the transport equation obtained either in Ref.~\cite{Son:2012zy} or in this paper
do not match exactly with the transport equation in Sec. IIB of Ref.~\cite{Son:2012zy}, which were obtained starting with a corrected form of the classical point-particle action,
with modified Poisson brackets. This starting point can be justified by performing a Foldy-Wouthuysen diagonalization of the quantum Dirac Hamiltonian, as seen in Ref.~\cite{Manuel:2014dza}.
However, the same exact form of the transport equation is not obtained if the starting point is a quantum field theory. Let us stress that in such a formulation one obtains the covariant form of
the chiral anomaly, as the chiral current is not defined by performing a functional derivative of an action, but from the equation obeyed by the current in the transport approach.

The question remains whether there can be more than one possible transport equation describing the same system equally well. The  Foldy-Wouthuysen diagonalization used in Ref.~\cite{Manuel:2014dza}
suggests that the starting quantum fields used there or those used in our OSEFT approach are not the same beyond the classical limit approximation. Thus, probably it is not so
surprising that one does not end up with the same exact form of the corresponding kinetic equations, while the two approaches give an equivalent description of the system.

Probably more surprising are the discrepancies we obtained from the results of Refs.~\cite{Hidaka:2016yjf,Hidaka:2017auj,Hidaka:2018ekt}, obtained from  massless QED, assuming
homogenous gauge field backgrounds. OSEFT only helps in
organizing the quantum field theory computation at large energies, as it has already been checked in the computation of Feynman diagrams at high $T$ \cite{Manuel:2016wqs,Carignano:2017ovz}. 
We cannot comment on the possible origin of these discrepancies, although it seems that the approach should also lead  to the consistent form of the chiral anomaly, rather than its covariant form,
as claimed in Ref.~\cite{Hidaka:2017auj}.

Let us, however, stress that discrepancies of our results with others published in the literature only appear
at order $n=2$  in the energy expansion both in the transport equation and the
current. Let us mention that since the chiral magnetic effect, as well as other chiral transport effects, appear already
at order $n=1$ our formulation gives the same description as that of other formulations (see Appendix~\ref{CME-app} for the computation of the chiral magnetic effect).

While in this paper we have focused our attention to the collisionless form of the transport equation, a much more challenging task is to derive the collision terms from OSEFT,
such that the Lorentz symmetry is respected, and the side-jumps are properly described. This will be the subject of a different project.

{\bf Acknowledgments:}
We are indebted to  J. Soto  for many discussions during the evolution of this project. 
We are also especially thankful to K. Landsteiner, for different discussions on the difference between the consistent and covariant forms of the chiral anomaly. We acknowledge interesting discussions with M. Beneke and T. Schaefer.
We have been supported by the MINECO (Spain) under Project No. FPA2016-81114-P.
This work was also supported by the COST Action CA15213 THOR. J.M.T.-R. was supported by the U.S. Department of Energy under Contract No. DE-FG-88ER40388. 
S. C. acknowledges financial support by the ``Fondazione Angelo Della Riccia."

\begin{appendix}

\section{Derivation of the $I_{\chi,\pm}$ functions}
\label{compute-Is}

We provide in this Appendix some details of the computation of the $I_{\chi,\pm}$ functions. We take here $e=1$ for simplicity.

We start from the equation of motion for quantum fields $\chi_v$,
\be
\left( {\cal O}_x^{(0)} +{\cal O}_x^{(1)} + {\cal O}_x^{(2)} \right) \chi_v(x) = 0 \,,
\ee
and similarly its Hermitian conjugate for $y$. By adding and subtracting
them, we can build equations for the two-point function.
For each piece, we isolate the different possible Dirac structures, so we
write
\be
{\cal O}^{(n)}_x = \left( \alpha_x^{(n)} + \beta^{(n)}_{x,\mu\nu}
\sigma_\perp^{\mu\nu} \right) \frac{\vts}{2} \ ;
\ee
then, taking the trace of \Eq{I-functions}, one gets

\be
\Tr(I_\pm^{(n)}) = \int d^4s e^{i k\cdot s} \left\lbrace
\left(\alpha_x^{(n)} \pm \alpha_y^{(n)*} \right)  \Tr \left[\frac{\vts}{2}
S_{E,v} (x,y)\right]  +
\left(\beta_{x,\mu\nu}^{(n)} \pm \beta_{y,\mu\nu}^{(n)*} \right) \Tr
\left[ \sigma^{\mu\nu} \frac{\vts}{2} S_{E,v}(x,y) \right] \right\rbrace \,,
\ee

For the $\alpha$ and $\beta$ coefficients, we find (after neglecting terms
of higher order in the gradient expansion like $\partial^X_\alpha
F_{\mu\nu}$ )
\begin{align}
\alpha^{(0)} & = i v \cdot D  \,,& \qquad
& \beta_{\mu\nu}^{(0)}  = 0 \,, \\
\alpha^{(1)} & = -\frac {1}{2E} D_\perp^2  \,, & \qquad
& \beta_{\mu\nu}^{(1)}   = -\frac {1}{4E}  F_{\mu\nu} \,, \\
\alpha^{(2)} & = \frac {1}{4E^2} (v^\alpha -\vt^\alpha) \Big(F_{\mu\alpha} D^\mu -i
D_\alpha D_\perp^2 \Big)   \,, & \qquad
& \beta_{\mu\nu}^{(2)}  = \frac{i}{4E^2} \left( F_{\mu\alpha} \vt^\alpha
D_\nu - \frac{1}{2} F_{\mu\nu} (v \cdot D - \vt \cdot D) \right)  \,.
\end{align}
 %
We now perform the change of variables to the center of mass and relative
  coordinates $X,s$. The recurring
combinations will be

\be
D_\alpha^x - (D_\alpha^y)^* = 2 \left(\partial_\alpha^s + i
A_\alpha(X)\right) \,, \qquad D_\alpha^x + (D_\alpha^y)^* =
\partial_\alpha^X + i s_\beta \partial^\beta A_\alpha(X) \,,
\ee
together with
\begin{align}
(D_\perp^x)^2 + ((D_\perp^y)^*)^2  & = 2 \Big( \partial_X \cdot \partial_s
+  i (\partial_X \cdot A(X) + A(X) \cdot \partial_X )
+ i s^\beta \partial^X_\beta A^\alpha(X) \left(\partial_\alpha^s + i
A_\alpha(X) \right) \Big) \,, \\
(D_\perp^x)^2 + ((D_\perp^y)^*)^2  & = 2 \Big( \partial_s^2 + 2 i A(X)
\cdot \partial_s -  A(X)^2 \Big)  \,,
\end{align}

We also use that
\be
\Tr \left[ \frac{\vts}{2} S_{E,v} \right] = 2 \sum_{\chi = \pm} G^\chi_{E,v} 
\,,\qquad\quad   \Tr \left[ \sigma^{\mu\nu} \frac{\vts}{2} S_{E,v}\right] = -
\sum_{\chi = \pm} \chi \epsilon^{\mu\nu\alpha\rho} \, \vt_\alpha J^\chi_{(E,v),\rho} 
\,,
\ee
where $G$ and $J$ are defined in \Eq{Wigner-current} and \Eq{basic2GF},
respectively.

For an example, we can work out the lowest order function. If here $k^\mu$ denotes the canonical momentum
then

\begin{align}
I_+^{(0)}  & = \int d^4s e^{i k\cdot s}   i v \cdot (D_x - D_y^*)
\sum_{\chi=\pm} 2 G^\chi_{E,v}(X,s)  e^{-i A s} 
\nonumber
\\
                & = \int d^4s e^{i k\cdot s}   i v \cdot  2 \left(-i k + i
A(X)\right) \sum_{\chi=\pm} 2 G^\chi_{E,v}(X,s) e^{-i A s} 
\nonumber
 \\
                & = 4 (v \cdot \bar{k}) \int d^4s e^{i \bar{k} \cdot s} 
\sum_{\chi=\pm}  G^\chi_{E,v}(X,s)  =  4 (v \cdot \bar{k})
\sum_{\chi=\pm}  G^\chi_{E,v}(X,\bar{k})
\end{align}
where now $\bar{k}^\mu = k^\mu -A^\mu$ is the canonical momentum.

\section{Chiral magnetic effect}
\label{CME-app}

In this Appendix we briefly show how from our formulation one can reproduce the chiral magnetic effect.
For this, we start from the current Eq.~(\ref{current}) and focus on its spatial components in the local rest frame $u^\mu = (1,0,0,0)$. After performing the $q_0$ integration, we get
\begin{equation}
j^i(X) = e \sum_{\chi =\pm} \int \frac{d^3q}{(2\pi)^3} \Big( \frac{q^i}{E_q} + \frac{S_\chi^{ij} \Delta_j}{E_q} - \frac{e}{2 E_q^2} S_\chi^{ij} F_{j\sigma} {\tilde v}^\sigma \Big) f^{\chi}(X,q)
\Big\vert_{q_0=E_q}\,,
\end{equation}
 with the dispersion relation in this frame given by
\begin{equation}
q_0 = E_q = \vert {\bf q} \vert \left(1 - e \chi \frac{{\bf B}\cdot{\hat{\bf{q}}}}{2 \aq^2} \right) \,.
\end{equation}

We now expand the distribution function using the dispersion relation, and assume we are in equilibrium so we can use the standard Fermi-Dirac expressions:
\begin{equation}
 f^{\chi}(X,q)\Big \vert_{q_0=E_q} = f^{\chi}(\aq) - e \chi \frac{{\bf B}\cdot{\hat{\bf{q}}}}{2 \aq} \frac{ df^{\chi}(\aq)}{d\aq} \,, \qquad f^{\chi}(\aq) = \frac{1}{1+ e^{(\aq-\mu_\chi)/T}} \,,
 \end{equation}
 this in turn eliminates all terms containing spatial derivatives, and keeping only the leading terms in $1/\aq$, we are left with
\begin{equation}
j^i(X) = e \sum_{\chi= \pm} \int \frac{d^3q}{(2\pi)^3} \Big[  \Big( \frac{q^i}{\aq}  - e \epsilon^{jkl}  \frac{S_\chi^{ij}}{\aq} B^l \frac{\partial}{\partial q^k} \Big) f^{\chi}(\aq) - e \frac{q^i}{\aq} \chi \frac{{\bf B}\cdot{\hat{\bf{q}}}}{2 \aq} \frac{ df^{\chi}(\aq)}{d\aq} \Big]\,.
\end{equation}

 After an integration by parts and performing angular integration, we finally arrive at
 \begin{equation}
 j^i(X) = - \frac{e^2}{4\pi^2} B^i \sum_{\chi =\pm} \chi \int d\aq \aq \frac{ df^{\chi}(\aq)}{d\aq} = e^2 \frac{\mu_5}{4\pi^2} B^i \,
\end{equation}
where $\mu_5 = \mu_1 - \mu_{-1}$, which is exactly the expected result for the chiral magnetic effect~\cite{Vilenkin:1980fu,Redlich:1984md,Fukushima:2008xe}; see also Ref.~\cite{Manuel:2013zaa}.

\end{appendix}

 \end{document}